\documentclass{jnmp03c}

\begin{document}

\renewcommand{\evenhead}{A~N~Leznov}
\renewcommand{\oddhead}{$UV$ Manifold and Integrable
Systems in Spaces of Arbitrary Dimension}

\setcounter{page}{1}

\thispagestyle{empty}

\FistPageHead{1}{\pageref{leznov-firstpage}--\pageref{leznov-lastpage}}{Letter}

\copyrightnote{2001}{A~N~Leznov}

\Name{$\pbf{UV}$ Manifold and Integrable Systems\\
 in Spaces  of Arbitrary Dimension}\label{leznov-firstpage}

\Author{A N LEZNOV}

\Adress{Research Center on Engeneering and Applied Science\\
Av. Universidad 1001 col. Chamilpa C.P. 62210, Cuernavaca Morelos Mexico\\[2mm]
Institute for High Energy Physics, 142284 Protvino, Moscow Region, Russia}

\Date{Received December 7, 1999; Revised March 10, 2000;
Accepted October 10, 2000}

\begin{abstract}
\noindent
The $2n$ dimensional manifold with two mutually commutative operators of
differentiation is introduced. Nontrivial multidimensional integrable systems
connected with arbitrary graded (semisimple) algebras are constructed.
The general solution of them is presented in explicit form.
\end{abstract}

\section{Introduction}

The success in the application of group-theoretical methods to the theory of
two-dimen\-sio\-nal integrable systems \cite{leznov:1,leznov:2} is not accidental. It
is connected with the
circumstance that the operations of multiplication of the group elements from
left and right are mutually commutative. It allows us  to associate with
two commutative differential operators $\frac{\partial}{\partial x}$,
$\frac{\partial}{\partial y}$ of two dimensional space
infinitesimal displacement generators dependent upon
one parameter from left and right  in such a way as to
solve trivially the representation of zero curvature type.

The foundation of the whole construction is the group valued function:
\be
K=M(x)N(y),\label{leznov:eq-1}
\ee
where the group valued elements $M$, $N$ are constructed by definite simple rules,
containing only operations in one dimensional spaces  $(x,y)$ respectively.

Many different attempts have been made to generalize this
construction to the multidimensional case by substituting instead
of $(x,y)$ in (\ref{leznov:eq-1}) some multidimensional functions
$x=\phi(x_1,\ldots,x_n)$, $y=\psi(y_1,\ldots,y_m)$ and considering
the consequences which follow from the associated multidimensional
$L$--$A$ pair formalism. However no interesting nontrivial
integrable system has been discovered in this way. Moreover the
solutions which arise in such a construction (in the concrete
examples  considered) are always only particular but not general.
The  reader can find the corresponding literature
in~\cite{leznov:RS}.

In the present paper we want exploit the following observation. It is always
possible to represent  general solutions of  integrable systems following from
(\ref{leznov:eq-1})  in  local form. Or in other words to obtain the general solutions of
such systems only the operation of differentiation is necessary (see \cite{leznov:LS}).
The operation of integration arises only on the middle step of calculations and
may be eliminated from the final result. This means that if we  have a
multidimensional space with two mutually commutative operators of
differentiation $D_1$, $D_2$ we will be able to use the two dimensional
construction to obtain nontrivial multidimensional integrable systems together
with their general solutions.

The aim of the present paper is in the explicit construction of multidimensional
manifolds with such properties. The paper is organized in the following way.
In Section~2 we discuss  general properties of the operators of differentiation
and obtain the equations they define. In Section~3 we find the solution of the
corresponding equations and in this realize the manifold with the
necessary properties. Section~4 is devoted to consideration of  concrete
examples of integrable systems on the multidimensional manifold constructed
above.
Comments and perspectives for further investigation are collected in Section~5.

\section{The main equations defining the manifold}

Suppose we have a $2n$ dimensional Euclidian space with a set of independent
coordinates $y_i$, $\bar y_i$. We assume the existence of two mutually commutative
generators of differentiation $D$, $\bar D$, which satisfy the basic relations:
\be
D \bar y=0,\qquad \bar D y=0\label{leznov:eq-2}
\ee
which are reminiscent of (anti) holomorphic functions in the theory of the
complex variables. So $D$ is a linear combination of space derivatives with
respect to the unbarred coordinates, $\bar D$ is the same with respect to the
barred ones:
\be
D=\frac{\partial}{\partial y_n}+\sum u^{\mu} \frac{\partial}{\partial y_{\mu}},
\qquad
\bar D=\frac{\partial}{\partial \bar y_n}+\sum v^{\nu} \frac{\partial}{\partial
\bar y_{\nu}}.\label{leznov:eq-3}
\ee
We choose the coefficient functions of differentiation with respect to $y_n$,
$\bar y_n$  to be equal to unity. This is an inessential restriction.

If we demand mutual commutativity of these generators $[D,\bar D]=0$, then as a
corollary of (\ref{leznov:eq-3}) we obtain the system of equations, which the coefficient
functions must satisfy:
\be
Dv^{\nu}\equiv v^{\nu}_{y_n}+\sum u^{\mu} v^{\nu}_{y_{\mu}}=0,\qquad
\bar Du^{\mu}=u^{\mu}_{\bar y_n}+\sum v^{\nu} u^{\mu}_{\bar y_{\nu}}=0.
\label{leznov:eq-4}
\ee
We will call (\ref{leznov:eq-4}) the
$uv$ system and the corresponding manifold the $UV$
one.

From (\ref{leznov:eq-4}) it follows that  an arbitrary function $\bar f$ of variables
$v$, $\bar y$ are annihilated by differentiation by $D$, all functions $f(u,y)$ by  $\bar D$.
In this sense we will speak about holomorphic and antiholomrphic functions.

The following proposition arises:

\medskip

\noindent {\bf Proposition 1.} {\it Each function $\bar f$
annihilated by  the operator $D$  is holomorphic; a function~$f$
annihilated by the operator $\bar D$ is antiholomorpthic.}

\medskip

Let us add  the equations $D \bar f=0$ and $\bar D f=0$ to the system of $(n-1)$
equations~(\ref{leznov:eq-4}). Then the $n$ sets of variables $(1,u)$, and $(1,v)$
respectively satisfy a linear system of $n$ algebraic equations the matrix of
which coincides with the Jacobian matrix (we consider the holomorphic case):
\be
J=\det_n \left|\begin{array}{cccc} v^1 & \dots & v^{n-1} & f \\
                    y_1 & \dots & y_{n-1} & y_n \end{array}\right|
\ee
which in the case of a non-zero solution of the linear system must
vanish. So Proposition~1 is proved.

As a corollary we obtain the following

\medskip

\noindent
{\bf Proposition 2.}
\be
\bar Dv^{\nu}=v^{\nu}_{\bar y_n}+\sum v^{\mu} v^{\nu}_{\bar y_{\mu}}=
V^{\nu}(v;\bar y),\qquad
Du^{\mu}=u^{\mu}_{y_n}+\sum u^{\nu} u^{\mu}_{y_{\nu}}=U^{\mu}(u;y).
\label{leznov:eq-BA}
\ee

Indeed operators $D$, $\bar D$ are commutative and so $\bar Dv$ is solution of the
same equation as $v$ satisfies. But each solution of third equation is a
holomorphic function, which proves  Proposition~2.

If we consider functions $U$, $V$ as given, then all operations of
differentiation applied to the functions $\Phi(u,v;y,\bar y)$ are
well defined.

As was mentioned in the introduction only these operations arise in the theory
of integrable systems constructed in the manner of (\ref{leznov:eq-1}). Thus for a
realization of the proposed program it is necessary to solve the system of
equations (\ref{leznov:eq-4}),
(\ref{leznov:eq-BA}) with the given holomorphic and antiholomorphic
$U$, $V$ functions.

\section{General solution of $\pbf{uv}$ system and realisation\\
 of the multidimensional $\pbf{UV}$ manifold}

Consider the system of equations defining implicitly $(n-1)$ unknown
functions $(\phi)$ in $(2n)$ dimensional space $(y,\bar y)$:
\be
Q^{\nu}(\phi;y)=P^{\nu}(\phi;\bar y)\label{leznov:eq-D}
\ee
with the convention that all Greek indices take values between $1$ and $(n-1)$.
The number of equations in (\ref{leznov:eq-D}) coincides with the number of unknown
functions $\phi^{\alpha}$. Each arbitrary function $Q$, $P$ depends on  $(2n-1)$
coordinates.

With the help of the usual rules of differentiation of  implicit
functions we find from (\ref{leznov:eq-D}): \be
\phi_y=(P_{\phi}-Q_{\phi})^{-1} Q_y,\qquad \phi_{\bar
y}=-(P_{\phi}-Q_{\phi})^ {-1}P_{\bar y}.\label{leznov:eq-DD} \ee
Let us assume, that between $n$ derivatives with respect to barred
and unbarred variables the following linear dependence takes
place:
\[
\sum^n_1 c_i \phi^{\alpha}_{y_i}=0,\qquad \sum^n_1 d_i \phi^{\alpha}_{\bar y_i}=0
\]
and analyse the corollaries following from these facts.

Assuming that $c_n\neq 0$, $d_n\neq 0$, dividing them into each
equation of the left and right sets respectively and introducing
the notation $u^{\alpha}=\frac{c_{\alpha}}{c_n}$,
$v^{\alpha}=\frac{d_{\alpha}}{d_n}$, we rewrite the last system in
the form: \be \phi^{\alpha}_{y_n}+\sum^{n-1}_1 u^{\nu}
\phi^{\alpha}_{y_{\nu}}=0,\qquad \phi^{\alpha}_{\bar
y_n}+\sum^{n-1}_1 v^{\nu} \phi^{\alpha}_{\bar y_{\nu}}=0.
\label{leznov:eq-MS} \ee Substituting values of the derivatives
from (\ref{leznov:eq-DD}) and multiplying result by the matrix
\mbox{$(P_{\phi}-Q_{\phi})$} on the left we obtain: \be
Q^{\alpha}_{y_n}+\sum^{n-1}_1 u^{\nu}
Q^{\alpha}_{y_{\nu}}=0,\qquad P^{\alpha}_{\bar y_n}+\sum^{n-1}_1
v^{\nu} P^{\alpha}_{\bar y_{\nu}}=0. \label{leznov:eq-D1} \ee From
the last equations it immediately follow that: \be u=-(Q_y)^{-1}
Q_{y_n},\qquad v=-(P_{\bar y})^{-1} P_{\bar y_n}.
\label{leznov:eq-UV} \ee We see that if we augment the initial
system (\ref{leznov:eq-D}), by $(n-1)$ vector functions $(u,v)$
defined by (\ref{leznov:eq-UV}) then the differential operators
$D$, $\bar D$ defined by (\ref{leznov:eq-3}) in connection with
(\ref{leznov:eq-MS}) annihilate every $\phi$ and therefore the
functions $Q$, $P$:
\be D \phi=\bar D \phi=D Q=D P=\bar D Q=\bar D
P=0. \label{leznov:eq-VIC} \ee This means that $D\bar f(\phi,\bar
y)=\bar D f(\phi, y)=0$. And as a direct corollary of this fact
$Dv=\bar D u=0$ and so the generators $D$, $\bar D$ constructed in
this way are mutually commutative.

Thus we have found the general solution of the $uv$ system and in such a way
realise the manifold with the properties postulated in the previous section.

It is possible to say (the solution of the $uv$ system is  general) that
this realisation is unique up to possible similarity transformations.

We present below the result of calculations of the functions $U$, $V$  using the
definition of $u$, $v$ functions (\ref{leznov:eq-UV}):
\[
\ba{l}
\ds U=D u=-Q_y^{-1}\left(D Q_{y_n}+\sum u^{\alpha} D Q_{y_{\alpha}}\right) ,
\vspace{3mm}\\
\ds V=\bar D v=-P_{\bar y}^{-1}\left(\bar D P_{\bar y_n}+\sum ^{\alpha} \bar D Q_{\bar
y_{\alpha}}\right).
\ea
\]

\section{Examples}

Below we would like to consider only two examples clarifying the situation. But
really the same is true with respect to each two dimensional system integrable
with the help of the formalism of graded algebras \cite{leznov:1,leznov:2}.

We want to emphasize that in all cases (particulary those considered below) the
$UV$ manifold by itself is defined by $2(n-1)$ arbitrary functions $Q$, $P$ each of
one of which  depends upon $2n-1$ independent arguments. All these functions
occur as coefficient functions (via operators of differentiation
$D$,
$\bar D$) in equations of multidimensional integrable systems. Naturally the
general solution  depends upon them. Apart from these functions the general
solution of an integrable system depends upon additional arbitrary functions
the number of which and their functional dependence has to be sufficient for
the statement of Cauchy or Gursat initial data problems.

\subsection{Multidimensional Liouville equation}

By this term we understand the equation:
\[
D\bar D \phi=\exp 2\phi.
\]
By direct calculation one can become convinced that its general solution has
the form:
\[
\exp\phi=\frac{(D c)^{1/2} (\bar D \bar c)^
{1/2}}{c+\bar c}.
\]
For instance in the four-dimensional case $(y_1,y_2;\bar y_1,\bar y_2)$:
\[
D\bar D=\bar D D=\frac{\partial^2}{\partial y_2 \partial \bar y_2}+
v\frac{\partial^2}{\partial y_2 \partial \bar y_1}+u\frac{\partial^2}
{\partial \bar y_2 \partial y_1}+uv\frac{\partial^2}{\partial y_1 \partial
\bar y_1}
\]
and
\[
u=-\frac{Q_{y_2}}{Q_{y_1}},\qquad u=-\frac{P_{\bar y_2}}{P_{\bar y_1}},\qquad
Q(\phi;y_1,y_2)=P(\phi;\bar y_1,\bar y_2).
\]
The functions $c(u;y)$, $\bar c(v;\bar y)$ are arbitrary holomrphic,
antiholomorphic functions.

\subsection{Multidimensional Toda system}

The equations of the Toda lattice in two dimensions have the form (one of the
many possible ones):
\[
x^i_{z,\bar z}=\exp \left(-x^{i-1}+2x^i-x^{i+1}\right),\qquad 1\leq i \leq n,\qquad
x_{-1}=x_{n+1}=0.
\]
The general solution may be represented as:
\[
\exp\left(-x^i\right)=\det_i \left\{ V^0 \right\},\quad
\exp(-x_0)=V^0,
\]
where matrix elements of the determinant matrix $V^0$ are as
follows:
\[
V^0_{i,j}=\frac{\partial^{i+j-2}}{\partial z^{i-1}\partial \bar z^{j-1}} V^0
\]
and the single function $V^0$ has the form:
\[
V^0=\frac{W_n(\theta,z) W_n(\bar \theta,\bar z)}{1+\sum\limits_1^n \theta_i\bar
\theta_i}.
\]
$W_n(\phi,x)$ is the determinant of the  Wronskian matrix with elements:
\[
W_{i,j}=\frac{\partial^{i-1}\theta^j}{\partial z^{j-1}}.
\]

To obtain the explicit general solution of multidimensional Toda lattice system:
\[
\bar D D=D\bar D x^i=\exp \left(-x^{i-1}+2x^i-x^{i+1}\right)
\]
only the following changes in all the corresponding formulae above are
necessary:
\[
\frac{\partial}{\partial z}\to D,\qquad \frac{\partial}{\partial \bar z}\to \bar D
\]
and consider arbitrary functions $\theta$, $\bar \theta$ as arbitrary holomorphic,
antiholomorphic functions on the manifold $UV$ of the corresponding dimension.

\section{Outlook}

The main results of the present paper are in the general solution of  the $uv$
system (\ref{leznov:eq-4}), a realisation of a multidimensional manifold with two
commutative operators of differentiation (Section~3) and
the integrable systems in the spaces of arbitrary dimensions constructed in
this framework (Section~4).

It turns out that with the help of $UV$ manifold it is possible to find
general solution of such interesting from the point of viw physical applications
systems as homogeneous
Complex Monge--Amph\'er and Bateman equations in the spaces of
arbitrary dimensions~\cite{leznov:combat}. This was achieved by reducing of the
definite kind the general solution of $uv$ system on subclasses in
which solution of it is functionally depends only on two arbitrary functions of
the necessary number of independent arguments.

It is obviousl that formalism of evolution type systems (integrable with
the help of the old inverse scattering method) remains
without any changings in the spaces of arbitrary dimension. The exactly
integrable systems considered above are the symmetries of evolution type
equations. The knowledge of the general solution of the first allow represent
in explicit form multisoliton solutions of the last.

We should like to finish this outlook with some speculative comments about the
possible application of the  proposed construction to the problems of the
physics.

If the manifold constructed would have any relation to the real four dimensional
world, then something similar to Einstein's General Relativity would occur.
Indeed, in both cases the general (geometrical) properties of the world be
determined by some fundamental physical objects, the metrical tensor $g_{ij}$ in
the case of General Relativity, Einstein's equations for which takes into account
the presence of all forms of matter and the equations $uv$ describing the
manifold $UV$ in the case considered above (which of course must be modified to
take into account all other physical fields). To be optimistic, it may  happen
that equations of General Relativity have a solution on the manifold of the
kind described above or something similar to it. Of course all this is only an
attractive speculation and only a deeper investigation of the problem may
clarify the situation.

\subsection*{ Acknowledgements}

The author gratefully thanks D~B~Fairlie during the common work with whom
on the problems of the Monge--Amph\'ere and the Bateman equation the idea of the manifold $UV$
was born, and A~B~Shabat for  discussions in the process of working on this
paper and important comments.

The author is indebted to the Center for Research on Engenering
and Applied Sciences (UAEM, Morelos, Mexico) for its hospitality and
Russian Foundation of Fundamental Researches (RFFI) GRANT N 98-01-00330 for
partial support.

\label{leznov-lastpage}

\newpage


\newpage


\begin{thebibliography}{99}
\small
\topsep0mm
\partopsep0mm
\parsep0mm
\itemsep0mm

\bibitem{leznov:1}
Leznov A N,
The Exactly Integrable Systems Connected with the Semisimple Algebras
of the Second Rank $A_2$, $B_2$, $C_2$, $G_2$,
{\it J. Nonlin. Math. Phys.}, 1999, V.6, 181--197; math-ph/9809012.

\bibitem{leznov:2}
Leznov  A N, Graded Lie Algebras, Representation Theory, Integrable Mappings and
Systems. Nonabelian Case, {\it Nucl. Phys. B}, 1999,
V.543, N~3, 652--672; math-ph/9810006.

\bibitem{leznov:LS} Leznov A N and Saveliev M V, Group Theoretical Methods for
Integration of Nonlinear Dynamical Systems,  Basel, 1992.

\bibitem{leznov:RS}
Razumov A V and Saveliev M V, Maximally Nonabelian Toda Systems,
hep-th/9612081.

\bibitem{leznov:combat}
Fairlie D B and Leznov A N,
The Complex Bateman Equation, {\it Lett. in Math. Phys.}, 1999, V.49, 213--216.

\end{thebibliography}
\end{document}